\documentclass[conference]{IEEEtran}
\IEEEoverridecommandlockouts
\usepackage{packages}
\def\BibTeX{{\rm B\kern-.05em{\sc i\kern-.025em b}\kern-.08em
    T\kern-.1667em\lower.7ex\hbox{E}\kern-.125emX}}
\begin{document}

\title{A Framework for Using LLMs for Repository Mining Studies in Empirical Software Engineering}

\author{
    \IEEEauthorblockN{
        Vincenzo De Martino\IEEEauthorrefmark{1},
        Joel Castaño\IEEEauthorrefmark{2},
        Fabio Palomba\IEEEauthorrefmark{1},
        Xavier Franch\IEEEauthorrefmark{2},
        Silverio Martínez-Fernández\IEEEauthorrefmark{2}
    }
    \IEEEauthorblockA{\IEEEauthorrefmark{1}Software Engineering (SeSa) Lab, University of Salerno, Italy\\
    \{vdemartino, fpalomba\}@unisa.it}
    \IEEEauthorblockA{\IEEEauthorrefmark{2}Universitat Politècnica de Catalunya, Spain\\
    \{joel.castano, xavier.franch, silverio.martinez\}@upc.edu}
}

\maketitle

\begin{abstract}
Context: The emergence of Large Language Models (LLMs) has significantly transformed Software Engineering (SE) by providing innovative methods for analyzing software repositories. 
Objectives: Our objective is to establish a practical framework for future SE researchers needing to enhance the data collection and dataset while conducting software repository mining studies using LLMs.
Method: This experience report shares insights from two previous repository mining studies, focusing on the methodologies used for creating, refining, and validating prompts that enhance the output of LLMs, particularly in the context of data collection in empirical studies. 
Results: Our research packages a framework, coined Prompt Refinement and Insights for Mining Empirical Software repositories (PRIMES), consisting of a checklist that can improve LLM usage performance, enhance output quality, and minimize errors through iterative processes and comparisons among different LLMs. We also emphasize the significance of reproducibility by implementing mechanisms for tracking model results. 
Conclusion: Our findings indicate that standardizing prompt engineering and using PRIMES can enhance the reliability and reproducibility of studies utilizing LLMs. Ultimately, this work calls for further research to address challenges like hallucinations, model biases, and cost-effectiveness in integrating LLMs into workflows.
\end{abstract}


\section{Introduction}
\label{sec:introduction}
The use of Large Language Models (LLMs) has brought about a breakthrough in software engineering (SE), enabling more efficient and straightforward approaches to complex tasks\cite{nam2024using,imran2024uncovering,hou2023large}. These models, trained on large datasets, have demonstrated their ability to assist in various SE activities \cite{white2024chatgpt}, including code generation \cite{tian2023chatgpt}, documentation generation \cite{della2024using}, and software project analysis \cite{tufano2024unveiling}. LLM-based  technologies provide software engineers with tools to automate tasks, enhancing code quality and accelerating their workflows. However, despite their increasing adoption in SE activities, the processes of prompt creation, improvement, and output validation remain unclear, which can lead to erroneous evaluations and results that do not meet stated goals \cite{fan2023large}. 
The work presented is a combination of lessons learned from previous work based empirical software engineering (ESE) experiences, seeking to address the challenges associated with the whole life cycle of using the LLM. We provide an experiential description of how to create, improve, and validate prompts through the use of single and multiple LLMs, focusing in particular on the analysis of software project repositories.

\workdone{This experience report aims to provide researchers and practitioners with a practical framework for using and advice on conducting mining repository studies with LLMs on software repositories such as GitHub and Hugging Face.}


\noindent \textbf{Context on our experiences.} Our work is grounded in two prior studies where we employed LLMs for mining software repositories. In the first study~\cite{de2024developers}, we analyzed 168 open-source ML projects on GitHub to assess the adoption of green architectural tactics for sustainable ML-enabled systems. We used LLM APIs to identify both documented and undocumented green tactics in Python files. In the second study \cite{castano2024change}, we conducted a longitudinal analysis of over 50,000 ML models on Hugging Face to understand how they evolve. By classifying commits and releases using LLMs---providing commit titles, edited files, and related information to the model---we uncovered patterns in model maintenance and development. These experiences highlighted the steps in prompt creation, improvement, and output validation when leveraging LLMs for software repository analysis.

\begin{figure}[ht]
    \centering
    \includegraphics[width=0.7\linewidth]{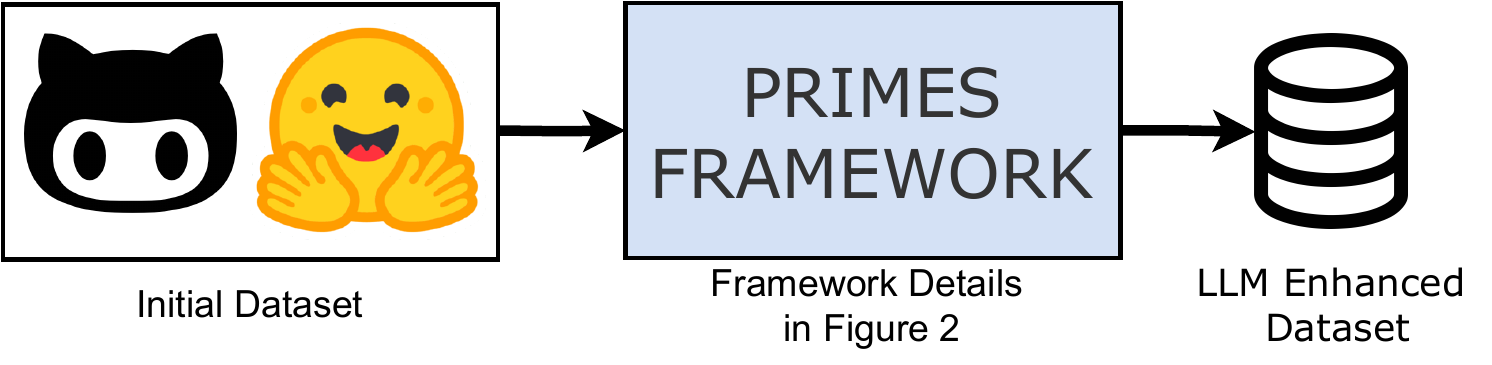}
    \caption{Enhancing a software repository using PRIMES for data collection}
    \label{dataset-enhancement}
\end{figure}

As depicted in Figure~\ref{dataset-enhancement}, our framework coined Prompt Refinement and Insights for Mining Empirical Software repositories (PRIMES), enables the transformation of an initial dataset into an enhanced dataset using LLMs, automating data collection and enrichment. This process streamlines the extraction of information from software repositories, facilitating more efficient empirical studies.

\section{Related work}
\label{sec:background}

\begin{figure*}[!ht]
    \centering
    \includegraphics[width=1\linewidth]{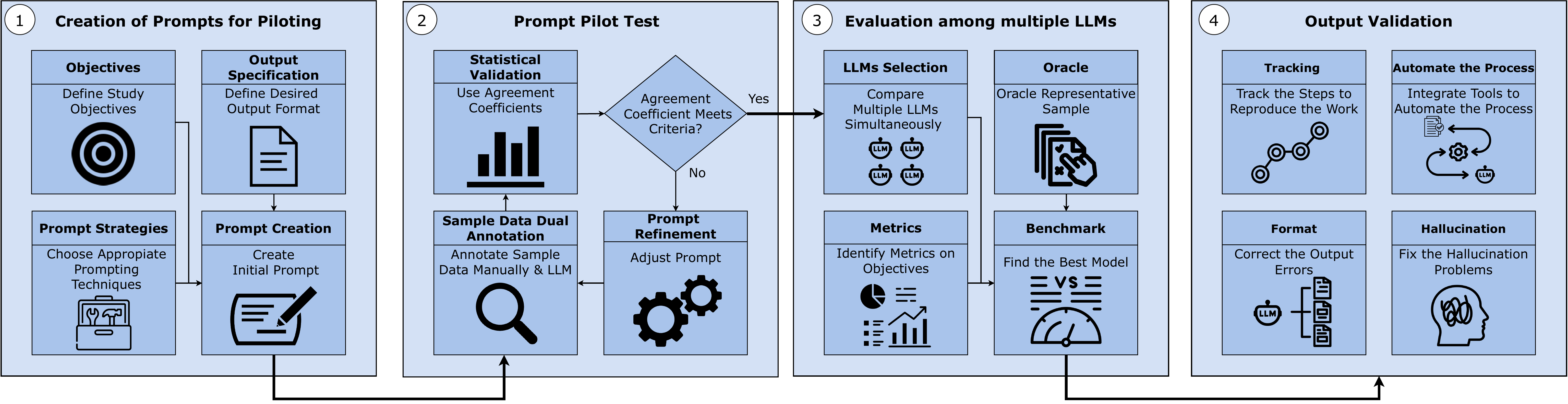}
    
    \caption{A preliminary, practical framework PRIMES for LLMs in automated data collection in mining repository studies, based on our experiences ~\cite{de2024developers, castano2024change}.}
    \label{fig:overview}
\end{figure*}
In the realm of SE, prior work has utilized LLMs for mining studies, showcasing their potential to automate and enhance various software development tasks. Silva et al. \cite{silva2024detecting} investigate the effectiveness of ChatGPT in detecting code smells in Java projects using a GitHub dataset containing four types of code smells classified by severity. The results show that detailed prompts, compared with generic prompts, significantly improve F-measure and can detect multiple critical code smells, suggesting the potential of LLMs to support software quality improvement through targeted prompt engineering. Shin et al. \cite{shin2023prompt} empirically assess the effectiveness of GPT-4 using various prompting strategies, revealing that while conversational prompts enhance performance, fine-tuned models often outperform LLMs in specific tasks. Their findings emphasize the importance of human feedback in refining prompts and suggest that further exploration of prompt engineering techniques is essential for optimizing LLM applications in automated SE. 
Fan et al.~\cite{fan2023large} examined the broader range of applications and open problems associated with LLMs in SE activities, highlighting the need for techniques that mitigate problems such as hallucinations and ensure the reliability of generated solutions. Zhang et al. \cite{zhang2024detecting} use LLMs and static analysis to find inconsistencies between Rust code and comments. Their method identified 160 mismatches across 13 open-source projects, showcasing LLMs' potential for automating documentation verification.



For the sake of our knowledge, our work advances the state of the art by creating the first, preliminary framework that enables the application of LLMs in software repository information extraction. This framework provides structured methodologies for creating prompts through iterative refinement and validation processes for the LLMs' outputs.


\section{Using LLMs for ESE: Methodological Insights}
\label{sec:insights}

Figure \ref{fig:overview} shows the proposed framework PRIMES, based on experiences from ~\cite{de2024developers, castano2024change}. PRIMES has four stages, mapping to the four subsections below and highlighted with numbers from {\large \ding{172}} to {\large \ding{175}}. Each stage consists of several steps, which are highlighted in \textit{italic} in the text below, and map to small boxes in dark blue in Figure \ref{fig:overview}. These steps can be seen as a preliminary checklist when designing repository mining studies using LLMs for automated data collection.

\subsection{Creation of Prompts for Piloting ({\large \ding{172}} in Figure \ref{fig:overview})}
Prompt engineering is an emerging field focused on creating and optimizing prompts to exploit LLMs across various applications and research domains effectively \cite{sahoo2024systematic}. It plays a key role in understanding and extending LLM functionality, enabling researchers and developers to design efficient techniques for interacting with LLMs, adapting their performance to specific needs, and integrating them seamlessly with other tools. Beyond constructing prompts, prompt engineering involves domain-specific skills and strategies, making it essential for leveraging LLM potential.

\textit{Defining study objectives:} Aligning the prompt with intended research outcomes, such as extracting specific information, improving data quality, or identifying SE practices, is essential. Clear objectives enable the development of targeted, context-specific prompts, ensuring consistent and goal-focused model outputs. The prompt should incorporate specific keywords, contextual details, and guiding questions to direct the model effectively.

\textit{Select appropriate prompt strategy:} LLMs follow instructions and generate context-dependent responses based on their training. Prompting techniques optimize results \cite{wei2021finetuned}, but their effectiveness varies by LLM and task complexity \cite{wang2024advanced}. Strategic prompt selection enhances performance, reduces costs, and supports sustainable LLM use in SE. Specifically:
\textit{(a)} one-shot prompting: providing a single example to guide the model's response, useful when data are limited;
\textit{(b)} few-shot prompting: supplying a few examples to help the model understand the task \cite{brown2020language}, valuable for meeting specific requirements without retraining;
\textit{(c)} chain-of-thought prompting: encouraging the model to break down complex tasks into intermediate reasoning steps \cite{wei2022chain}, improving performance in tasks requiring logical reasoning or multi-step problem solving;
and \textit{(d)} structured prompting: formatting the prompt to impose constraints or guide the model toward specific response formats, ensuring consistency and facilitating information retrieval.

Another key aspect is \textit{structuring the output} to facilitate analysis. The prompt should guide the LLM in generating accurate content organized in formats that support easy analysis, such as bullet points, tables, or categorized sections. Proper structuring reduces ambiguity, enhances clarity, and simplifies data extraction and analysis, leading to more reliable insights.

\textit{Initial Prompt Development:} Crafting an initial prompt that clearly communicates the task to the LLM is crucial. The prompt should include:
\textit{(a)} task description: a concise explanation of what is expected from the LLM;
\textit{(b)} contextual information: necessary background information or definitions relevant to the task;
and \textit{(c)} output format specification: instructions on how the LLM should structure its output (e.g., JSON, lists, tables).
The goal is to provide sufficient guidance to the LLM without overwhelming it with unnecessary details, ensuring outputs align with empirical research objectives.

\lessonlearned{Structured prompting and clear objectives are essential for effective LLM use. Defining objectives aligns prompts with research goals, while structured formats enhance clarity, consistency, and analysis, simplifying prompt development.}

\subsection{Prompt Pilot Test: Validation and Iterative Refinement of the Prompt on a Single LLM ({\large \ding{173}})} \label{sec:pilot}

This section describes a method for validating and improving prompts when using a single LLM.

\textit{Iterative Refinement Process:} The process involves the following steps:

\begin{enumerate}
    \item  \textit{Sample Data Dual Annotation}: Select a sample of data items. Annotate this data both using the LLM and manually by a human annotator, independently of each other.
    \item \textit{Statistical Validation}: Perform statistical validation by calculating agreement coefficients (e.g., Cohen's kappa~\cite{cohen1960coefficient}) to quantitatively assess the alignment between the LLM's outputs and the expected results.

    \item \textit{Criteria Assessment}: Check if the agreement coefficient meets a predefined criterion (e.g., Cohen's kappa $> 0.9$). If the agreement is acceptable, proceed to use the prompt; if not, continue to the next step.

    \item \textit{Prompt Refinement}: If the agreement is below the acceptable threshold, refine the prompt to address any identified issues. Refinement strategies may include:
    \textit{(a)} simplifying language: using clear and straightforward language to improve comprehension;
    \textit{(b)} clarifying instructions: rewriting parts of the prompt to eliminate ambiguity;
    \textit{(c)} providing examples: including examples to guide the LLM towards the desired output;
    and \textit{(d)}, adjusting level of detail: modifying the amount of context provided.
    \item \textit{Repeat Process:} Repeat the process starting from step 1 with the refined prompt until the statistical validation meets the criteria.

\end{enumerate}

During this assessment, we observed that incorrect outputs were often due to prompt complexity, which resulted in confusing responses. 
Using simple and clear prompts helps prevent overwhelming the LLM and improves its ability to generate accurate responses.

\lessonlearned{For effective software repository analysis with LLMs, simple and concise instructions should be used. According to our experience, wrong outputs may be due to prompt complexity, which results in confusing outputs. 
Clear guidance helps the model understand the task better, leading to more precise results.}

\exampleapp{In our previous study~\cite{castano2024change}, we applied this iterative refinement process to improve a prompt that classified commits according to a maintenance taxonomy adapted for machine learning projects. We began by selecting a sample of commit messages and their associated code changes. Both the LLM and human annotators independently classified these commits. Initially, the agreement measured by Cohen's kappa was below our acceptable threshold. We identified that the prompt was too complex and lacked clarity. By simplifying the language and providing specific examples for each classification category within the prompt, we enhanced the LLM's understanding. 

After refining the prompt, we repeated the validation process, and the agreement coefficient increased to meet our criterion ($> 0.9$), demonstrating that the prompt effectively guided the LLM to produce reliable classifications. This iterative process continues until the LLM's outputs consistently meet the desired standards when tested on the sample data.}

\subsection{Evaluation among multiple LLMs ({\large \ding{174}})} 

\textit{LLMs Selection:} The development of new LLMs is steadily increasing \cite{zhao2023survey}, and to assess the suitability of the best LLM for a given ESE task, it is important to conduct a comparative analysis between models. This provides insight into individual strengths and weaknesses, particularly in relation to the specific task at hand. This comparison is essential because these models were developed using different architectures and training data, which may impact task performance. 
This comparison reveals the generalizability of prompts and assesses which LLMs are best suited for specific research objectives.

\textit{Oracle creation:} An essential step in evaluating LLM results is the use of an oracle \cite{de2024developers}, which provides a reliable reference for validating model performance against a known ground truth. The oracle assesses the accuracy and reliability of LLM outputs, reduces ambiguity, and supports iterative improvement in prompting strategies. A well-defined oracle enhances reproducibility, allowing consistent evaluations across studies and applications.
\textit{(a)} Oracle construction: When constructing an oracle, an initial assessment on a sufficiently large, representative data subset is crucial to minimize bias and accurately reflect the larger dataset \cite{baltes2022sampling}. A sample size calculator can help estimate the required sample size for statistically significant conclusions\footnote{For instance: \url{https://www.qualtrics.com/blog/calculating-sample-size/}}.  
This approach allows for reliable and valid results, thus improving the robustness of the validation process and enabling more valid assessments of model performance. 
    
\textit{(b)} Expertise in oracle construction: Accurate classification of data within oracle requires domain expertise. Those constructing the oracle must have experience with the topic, whether it is the implementation of the code or the domain-specific tasks on which the LLM is evaluated. This initial phase could focus on well-documented and well-understood examples in the literature to maintain the oracle's validity. Without sufficient expertise, the oracle may be inaccurate, leading to incorrect evaluations of the model's performance. 

\exampleapp{In our study \cite{de2024developers}, we used multiple LLMs to evaluate their effectiveness in extraction studies, focusing on identifying green architectural tactics. In particular, we compared two models: GPT-4o from OpenAI \cite{gptkey} and Claude 3 Haiku from Anthropic \cite{haikukey}. The following metrics were used to determine which LLM was most appropriate to apply to the entire sample:

\begin{itemize}
    \item Classification accuracy: The primary metric used to evaluate the models was overall classification accuracy, measured using the previously described oracle. Accuracy offered a quantitative measure of each model's ability to classify. 
    \item Interpretability of explanations: In addition to raw accuracy, we also evaluated the quality of model explanations. The depth of these explanations was important for understanding and trusting model decision making, making interpretability a key secondary metric.
    \item Cost of using LLMs: Given the large file size for each ML project, the input and output cost of each API call became an important metric. In our case, we calculated the cost of analyzing large code bases using the price based on one million tokens in input and output of each LLM. This comparison is useful in making an economic choice for analyzing a large number of files, particularly in large-scale studies where cost utilization is a critical factor.
\end{itemize}

The models show different capabilities in terms of both their ability to identify tactics found in the literature and their ability to identify new tactics. Specifically, GPT-4o achieved a classification accuracy of 95.58\%, while Claude 3 Haiku achieved 97.91\%. This helped us determine which was the most accurate.
Both models effectively classify tactics from the literature with varying accuracy but fail to identify new ones. The GPT-4o model did not find any new green tactics, while Claude 3 Haiku managed to identify some additional sustainable tactics from our original catalog, though it occasionally repeated known ones.
These considerations allowed us to evaluate the performance of multiple LLMs, enabling us to select the most appropriate model for our study based on a combination of criteria. However, it is important to note that future studies could consider additional metrics depending on their specific research objectives.}

Limiting the classification to known examples ensures that the oracle is built on a solid foundation. In our approach \cite{de2024developers}, we limited manual classification to green tactics already documented in the literature \cite{jarvenpaa2024synthesis}, using well-defined criteria and examples that facilitated accurate identification. All new or undocumented tactics were set aside for later analysis when the LLM could help us discover them.
By building the oracle through these processes, the prompt used for the LLM can be refined iteratively, allowing for continuous improvement and validation of model results.

\textit{Benchmark:} Establishing a benchmark is crucial for fairly evaluating a prompt's performance across multiple LLMs. \revised{A poorly designed benchmark can lead to overconfidence in the LLM outputs, potentially creating unrealistic scenarios that fail to reveal limitations.}

\revised{Design a benchmark with tests to assess generalization, including false positives to test accuracy and robustness, and ensure task diversity to assess adaptability. This should be guided by a \textit{comprehensive set of metrics} to compare performance and align assessments with specific goals.} 

Additionally, an important aspect is the execution of multiple tasks simultaneously. Evaluating models involves assessing their capacity to manage various tasks, as using a single prompt for multiple tasks can lead to uneven performance and incomplete results. Monitoring task-specific performance reveals the models' generalization capabilities, helping identify strengths and areas for improvement in multitasking.

\lessonlearned{To effectively use of LLMs for data collection, it is crucial to establish a systematic process. Creating statistical samples and smaller preliminary studies to compare multiple LLMs are essential to ensure (or optimize) the use of the right LLM in the right way.}


\subsection{Output Validation ({\large \ding{175}})}
The output validation is essential for ensuring the data collection and analysis of LLM results, facilitating replication, and enabling future studies to build upon the work. 

Several handling issues may arise when evaluating model outputs. When validation uncovers issues such as hallucinations, duplications, or formatting errors, these have to be systematically addressed.

\textit{Correct Output Format:} Correct the output format---whether text, tables, or images---in the prompt is essential for managing how outputs are saved and analyzed. Clear specifications help structure model outputs, making it easier to derive insights for quantitative and qualitative analyses. This structured approach supports informed decision-making, especially for evaluating patterns or conducting manual reviews. 

\textit{(a)} Detecting duplications: Duplications arise when the LLM unnecessarily repeats information within or across outputs, leading to skewed analyses and incorrect conclusions. Use text similarity algorithms or hash-based methods to compare outputs and identify repeated content. Removing or consolidating duplications ensures each data point is unique.

\textit{(b)} Ensuring Correct Formatting: Consistent output formatting is essential for data processing and analysis. Formatting errors can hinder parsing routines and introduce errors into the analysis pipeline. Define the expected output structure explicitly and use validation scripts to check each output against this structure. Correcting these issues may involve adjusting the prompt for stricter adherence to the format or applying post-processing steps to standardize the outputs.

\textit{Checking for Hallucinations:} Hallucinations occur when the LLM generates content not present in the input data or contradicts facts, introducing false information into the study. To detect them, cross-reference the LLM's outputs with the original data to ensure all content is grounded in the provided information. Automated scripts can flag outputs containing unexpected terms or irrelevant concepts. Incorporating domain knowledge and predefined constraints helps recognize and eliminate hallucinated content.

\textit{Tracking Information:} Tracking where specific information appears is essential for reproducibility, allowing researchers to validate findings and understand the context of generated outputs. This is critical for building confidence in research outcomes, especially when working with complex models like LLMs and large datasets. Keeping detailed records of model outputs aids in analyzing the source and verifying the veracity of the information. For example, a script can traverse repository files, save project locations, and collect LLM outputs, generating a CSV file for each project that captures file names and corresponding outputs.

\textit{Automating the Process:} Automating the validation process is critical to ensure the accuracy and integrity of LLM outputs. This involves systematically identifying and addressing issues such as hallucinations, duplications, and formatting errors that can affect data quality and analysis. For instance, outputs expected in structured formats like JSON can be automatically validated using tools that parse data and report schema deviations, ensuring they meet predefined requirements.
An effective approach is to use data testing frameworks like ``Great Expectations''~\cite{greatexpectationsHomeGreat}, which facilitate automated and repeatable data quality checks by defining expectations or rules the data should satisfy.
Automating the validation process enhances efficiency and reduces human error. Tools like ``Great Expectations'' enable the creation of validation suites that can be executed programmatically, including checks for data types, value ranges, pattern matching, and schema conformity. Integrating an automated validation process into the workflow allows early detection of issues and prompt feedback for prompt adjustments or re-generation of outputs.

\lessonlearned{Ensuring reproducibility is crucial to validate research findings and build confidence in the results. It is important to trace the origins of specific model results, as inconsistencies and hallucinations in LLMs' responses can compromise reliability. Implementing tracking mechanisms and conducting validation checks can help maintain the integrity of the process.}
\section{Discussions and Outlook}
\label{sec:discussion}
This section examines the implications of using LLMs to analyze software repositories. Our findings guide future research directions.

\textbf{Limitations and Challenges of Using LLMs in ESE:} Despite the promising potential of LLMs in software repository mining, several limitations and challenges must be acknowledged. LLMs can generate hallucinations—fabricated or inaccurate information—that may lead to incorrect conclusions. They may also inherit biases from their training data, impacting fairness. The black-box nature of these models makes it difficult to interpret outputs and understand their reasoning. Additionally, the stochastic nature of LLMs complicates reproducibility, and practical issues like API limitations and costs can hinder large-scale studies.


To address these challenges, we implemented mitigation strategies, including rigorous validation procedures using expert-based oracles to evaluate LLM outputs and reduce inaccuracies. We assessed output reliability with statistical measures like Cohen's kappa and compared multiple LLMs based on metrics such as classification accuracy, interpretability, and cost-effectiveness to find the most suitable model. We acknowledge that model biases and the black-box nature of LLMs still require further research and attention. Moreover, challenges such as ensuring data privacy and addressing ethical concerns when using proprietary LLMs necessitate careful consideration. 
\revised{Another critical but often overlooked challenge is the environmental cost impact of training and deploying large-scale LLMs. Incorporating this dimension into the assessment framework ensures that trade-offs are aligned with sustainable AI principles.} Additionally, integrating LLMs effectively into existing SE workflows presents practical difficulties. Future research should explore these issues to fully harness the potential of LLMs in ESE.

\textbf{Importance of Effective Prompt Engineering:} Besides prompt engineering, which is an essential component of the work, we may want to emphasize the importance of a structured process to the use of LLMs. Well-defined and validated prompts are essential for generating accurate and relevant outputs. Previous studies \cite{silva2024detecting, shin2023prompt} focused only on the first step of the proposed framework in Figure \ref{fig:overview}, limiting their potential for improvement. LLMs should undergo a preliminary evaluation process that includes prompt engineering, refinement, accuracy assessment, and testing multiple models. For example, Silva et al. \cite{silva2024detecting} only used ChatGPT to find code smells in a Java code dataset. However, comparing multiple LLMs with the same prompt could yield better results due to differing architectures and training data. As such, this experience report provides a concrete framework for reporting the set of activities we have found useful in conducting research with LLMs. 


\textbf{Standardization of LLMs Guidelines:} There is a need to standardize both the reporting of results and the development of prompt engineering practices within mining studies using LLMs. Establishing clear guidelines for consistent reporting will improve comparability across studies, resulting in more robust analyses and better-informed decisions \cite{codabuxteaching}. Standardizing prompt engineering practices will provide a systematic approach to creating and refining prompts, leading to more effective and reliable results from LLMs. To facilitate this, we advocate for the integration of an LLM usage framework PRIMES into existing empirical standards. Incorporating such guidelines will help standardize methodologies and reporting practices when employing LLMs in mining studies.

\textbf{Enhancing Reproducibility:} Our research highlights the importance of reproducibility when working with LLMs in software repository mining. By systematically tracking model outputs and their sources, we have developed a framework that ensures the transparency and traceability of our findings. Prioritizing reproducibility helps validate results and provides a solid foundation for future research efforts. However, challenges such as model hallucinations and biases need to be acknowledged, as they can undermine trust in LLM outputs.

\textbf{Extension and Adaptability of the Proposed Framework:} While our study focused on particular aspects of software repository mining, the methods and insights gained can be adapted to various contexts. Future research could investigate the relevance of our results across different software domains to assess the generalizability of our framework PRIMES. Additionally, incorporating feedback from subsequent studies and user experiences will enable iterative improvements, enhancing the framework's adaptability to diverse analytical scenarios, such as code quality assessment and automated testing. One potential direction includes the creation and extension of a metrics catalog to benchmark multiple LLMs \revised{and consider their trade-off.} Exploring the integration of emerging technologies can further enrich the framework's capabilities, making it a robust resource for researchers and practitioners.


In conclusion, our findings contribute to advancing the use of LLMs in software repository mining studies. By sharing our experiences and a preliminary framework, we aim to both: (a) promote the right usage of LLMs for dataset enhancement (for automated and efficient data collection, requiring less manual work), considering monetary expenditure, time costs, and potential impact on the environment \cite{wang2024advanced}; and, (b) foster a collaborative environment in the ESE community encouraging continuous improvement and innovation in the acknowledged the limitations and challenges that pave the future work.

\section*{Acknowledgment}
This work has been partially funded by the Spanish research project GAISSA (TED2021-130923B-I00 by MCIN/ AEI/10.13039/501100011033). This work has been partially supported by the European Union - NextGenerationEU through the Italian Ministry of University and Research, Projects PRIN 2022 ``QualAI: Continuous Quality Improvement of AI-based Systems'' (grant n. 2022B3BP5S , CUP: H53D23003510006). 
\balance
\bibliographystyle{IEEEtran}
\bibliography{bib}

\begin{thebibliography}{10}
\providecommand{\url}[1]{#1}
\csname url@samestyle\endcsname
\providecommand{\newblock}{\relax}
\providecommand{\bibinfo}[2]{#2}
\providecommand{\BIBentrySTDinterwordspacing}{\spaceskip=0pt\relax}
\providecommand{\BIBentryALTinterwordstretchfactor}{4}
\providecommand{\BIBentryALTinterwordspacing}{\spaceskip=\fontdimen2\font plus
\BIBentryALTinterwordstretchfactor\fontdimen3\font minus \fontdimen4\font\relax}
\providecommand{\BIBforeignlanguage}[2]{{%
\expandafter\ifx\csname l@#1\endcsname\relax
\typeout{** WARNING: IEEEtran.bst: No hyphenation pattern has been}%
\typeout{** loaded for the language `#1'. Using the pattern for}%
\typeout{** the default language instead.}%
\else
\language=\csname l@#1\endcsname
\fi
#2}}
\providecommand{\BIBdecl}{\relax}
\BIBdecl

\bibitem{nam2024using}
D.~Nam, A.~Macvean, V.~Hellendoorn, B.~Vasilescu, and B.~Myers, ``Using an llm to help with code understanding,'' in \emph{Proceedings of the IEEE/ACM 46th International Conference on Software Engineering}, 2024, pp. 1--13.

\bibitem{imran2024uncovering}
M.~M. Imran, P.~Chatterjee, and K.~Damevski, ``Uncovering the causes of emotions in software developer communication using zero-shot llms,'' in \emph{Proceedings of the IEEE/ACM 46th International Conference on Software Engineering}, 2024, pp. 1--13.

\bibitem{hou2023large}
X.~Hou, Y.~Zhao, Y.~Liu, Z.~Yang, K.~Wang, L.~Li, X.~Luo, D.~Lo, J.~Grundy, and H.~Wang, ``Large language models for software engineering: A systematic literature review,'' \emph{ACM Transactions on Software Engineering and Methodology}, 2023.

\bibitem{white2024chatgpt}
J.~White, S.~Hays, Q.~Fu, J.~Spencer-Smith, and D.~C. Schmidt, ``Chatgpt prompt patterns for improving code quality, refactoring, requirements elicitation, and software design,'' in \emph{Generative AI for Effective Software Development}.\hskip 1em plus 0.5em minus 0.4em\relax Springer, 2024, pp. 71--108.

\bibitem{tian2023chatgpt}
H.~Tian, W.~Lu, T.~O. Li, X.~Tang, S.-C. Cheung, J.~Klein, and T.~F. Bissyand{\'e}, ``Is chatgpt the ultimate programming assistant--how far is it?'' \emph{arXiv preprint arXiv:2304.11938}, 2023.

\bibitem{della2024using}
A.~Della~Porta, V.~De~Martino, G.~Recupito, C.~Iemmino, G.~Catolino, D.~Di~Nucci, and F.~Palomba, ``Using large language models to support software engineering documentation in waterfall life cycles: Are we there yet?'' 2024.

\bibitem{tufano2024unveiling}
R.~Tufano, A.~Mastropaolo, F.~Pepe, O.~Dabi{\'c}, M.~Di~Penta, and G.~Bavota, ``Unveiling chatgpt’s usage in open source projects: A mining-based study,'' in \emph{2024 IEEE/ACM 21st International Conference on Mining Software Repositories (MSR)}.\hskip 1em plus 0.5em minus 0.4em\relax IEEE, 2024, pp. 571--583.

\bibitem{fan2023large}
A.~Fan, B.~Gokkaya, M.~Harman, M.~Lyubarskiy, S.~Sengupta, S.~Yoo, and J.~M. Zhang, ``Large language models for software engineering: Survey and open problems,'' in \emph{2023 IEEE/ACM International Conference on Software Engineering: Future of Software Engineering (ICSE-FoSE)}.\hskip 1em plus 0.5em minus 0.4em\relax IEEE, 2023, pp. 31--53.

\bibitem{de2024developers}
V.~De~Martino, S.~Mart{\'\i}nez-Fern{\'a}ndez, and F.~Palomba, ``Do developers adopt green architectural tactics for ml-enabled systems? a mining software repository study,'' \emph{arXiv preprint arXiv:2410.06708}, 2024.

\bibitem{castano2024change}
J.~Casta{\~n}o, R.~Caba{\~n}as, A.~Salmer{\'o}n, D.~Lo, and S.~Mart{\'\i}nez-Fern{\'a}ndez, ``How do machine learning models change?'' 2024.

\bibitem{silva2024detecting}
L.~L. Silva, J.~Silva, J.~E. Montandon, M.~Andrade, and M.~T. Valente, ``Detecting code smells using chatgpt: Initial insights,'' in \emph{Proceedings of the 18th ACM/IEEE International Symposium on Empirical Software Engineering and Measurement}, 2024, pp. 400--406.

\bibitem{shin2023prompt}
J.~Shin, C.~Tang, T.~Mohati, M.~Nayebi, S.~Wang, and H.~Hemmati, ``Prompt engineering or fine tuning: An empirical assessment of large language models in automated software engineering tasks,'' \emph{arXiv preprint arXiv:2310.10508}, 2023.

\bibitem{zhang2024detecting}
Y.~Zhang, ``Detecting code comment inconsistencies using llm and program analysis,'' in \emph{Companion Proceedings of the 32nd ACM International Conference on the Foundations of Software Engineering}, 2024, pp. 683--685.

\bibitem{sahoo2024systematic}
P.~Sahoo, A.~K. Singh, S.~Saha, V.~Jain, S.~Mondal, and A.~Chadha, ``A systematic survey of prompt engineering in large language models: Techniques and applications,'' \emph{arXiv preprint arXiv:2402.07927}, 2024.

\bibitem{wei2021finetuned}
J.~Wei, M.~Bosma, V.~Y. Zhao, K.~Guu, A.~W. Yu, B.~Lester, N.~Du, A.~M. Dai, and Q.~V. Le, ``Finetuned language models are zero-shot learners,'' \emph{arXiv preprint arXiv:2109.01652}, 2021.

\bibitem{wang2024advanced}
G.~Wang, Z.~Sun, Z.~Gong, S.~Ye, Y.~Chen, Y.~Zhao, Q.~Liang, and D.~Hao, ``Do advanced language models eliminate the need for prompt engineering in software engineering?'' \emph{arXiv preprint arXiv:2411.02093}, 2024.

\bibitem{brown2020language}
T.~B. Brown, ``Language models are few-shot learners,'' \emph{arXiv preprint arXiv:2005.14165}, 2020.

\bibitem{wei2022chain}
J.~Wei, X.~Wang, D.~Schuurmans, M.~Bosma, F.~Xia, E.~Chi, Q.~V. Le, D.~Zhou \emph{et~al.}, ``Chain-of-thought prompting elicits reasoning in large language models,'' \emph{Advances in neural information processing systems}, vol.~35, pp. 24\,824--24\,837, 2022.

\bibitem{cohen1960coefficient}
J.~Cohen, ``A coefficient of agreement for nominal scales,'' \emph{Educational and psychological measurement}, vol.~20, no.~1, pp. 37--46, 1960.

\bibitem{zhao2023survey}
W.~X. Zhao, K.~Zhou, J.~Li, T.~Tang, X.~Wang, Y.~Hou, Y.~Min, B.~Zhang, J.~Zhang, Z.~Dong \emph{et~al.}, ``A survey of large language models,'' \emph{arXiv preprint arXiv:2303.18223}, 2023.

\bibitem{baltes2022sampling}
S.~Baltes and P.~Ralph, ``Sampling in software engineering research: A critical review and guidelines,'' \emph{Empirical Software Engineering}, vol.~27, no.~4, p.~94, 2022.

\bibitem{gptkey}
``Chatgpt api keys,'' \url{https://openai.com/index/openai-api/}, 2024.

\bibitem{haikukey}
``Getting started - anthropic,'' \url{https://docs.anthropic.com/en/api/getting-started}, 2024.

\bibitem{jarvenpaa2024synthesis}
H.~J{\"a}rvenp{\"a}{\"a}, P.~Lago, J.~Bogner, G.~Lewis, H.~Muccini, and I.~Ozkaya, ``A synthesis of green architectural tactics for ml-enabled systems,'' in \emph{Proceedings of the 46th International Conference on Software Engineering: Software Engineering in Society}, 2024, pp. 130--141.

\bibitem{greatexpectationsHomeGreat}
``{H}ome | {G}reat {E}xpectations --- docs.greatexpectations.io,'' \url{https://docs.greatexpectations.io/docs/home/}, [Accessed 29-10-2024].

\bibitem{codabuxteaching}
Z.~Codabux, F.~Fard, R.~Verdecchia, F.~Palomba, D.~Di~Nucci, and G.~Recupito, ``Teaching mining software repositories.''

\end{thebibliography}
\end{document}